\documentstyle[12pt,version2,aps]{revtex}
\begin{document}
%
 
\begin{center}
{\large\bf ENHANCEMENT of
 PERSISTENT CURRENT ON MULTICHANNEL RING$^*$}

\vspace{.5 in}
{\sc F.V. Kusmartsev}
 \vspace{.3 in}

{\it L.D. Landau Institute for Theoretical Physics,
Moscow,117940, Russia }
\vspace{.3 in}

{\it NORDITA, Blegdamsvej 17, DK-2100, Kopenhagen, Denmark\par
Feo@NORDITA.DK}
\end{center}
\vspace{.9 in}
\date{}

\begin{abstract}

We describe a "train" effect which may exist in small metallic and
semiconductor rings and might be associated with the long-standing
problem  of the persistent current enhancement.
 The current is associated with
the cooperative motion of  $N$ electrons constituing
the $N$-electron  "train". The train arises  via an excitation
or a backflow  of spin waves or  of interchannel
 charge density sound modes.
The impurities and defects have a little  effect on the "train" current.
The reason is that the "train" current is 
associated with a small momentum transfer 
$\Delta k=k_{free}/N$, which is much smaller than the momentum transfer of
the free electron current equal to $k_{free}=2\pi/L$.
For an illustration of the "train" effect  we have 
calculated the persistent current in the framework of 
 the Bethe
ansatz solutions for Hubbard  model. The fractional $M/N$ periodicities of the persistent current serve as an  indication of the electron "train".
\end{abstract}
  
\vfill
\eject

There is a long-standing problem of the strong persistent current
observed in metallic\cite{22,21} and semiconductor rings\cite{33}.
The present existing theories describe reasonably
well the observed 
periodicities  but fail to explain the strength
of the current\cite{Land,Alte0,Bouc,Ambe0,Mull,Kus,Weis,Bouz}. 
In real rings used in experiments\cite{22,21,33} there
 are always impurities and
defects. However the value of the current is 
similar to the one in  systems without   defects.
 The calculation of the current in cases when these defects
have been
 taken into account by the perturbation theory
  gives the  current value
which is about  two or (for metallic rings) 
 four order of magnitude smaller than the observed one.

Thus,  the experiments
indicate that  impurities and  defects have no much influence
on the electrons. The persistent current
is not suppressed by defects as it is used to be in 1D or 2D metals\cite{Alte0}.
If we take into account the impurities on a single channel ring with free electrons, the disorder
 induces a repulsion between energy levels 
   in the regions of the level-crossing\cite{Bouc}. The level repulsion 
 has a simple explanation based on the picture of single electron motion
in a periodic potential. The motion of an 
electron on the ring at some configuration of the disorder may be
thought as the motion of an electron in the periodic potential
with the period equal to the circumference of the ring $L$.
But for the motion of a particle in a periodic potential,
according to the Bloch theorem, there are  bands and gaps occuring.
The values of the gaps are determined by the values of the barriers,
created by  disorder, which are proportional to the average
amplitude of the
disorder $W$.
The widths of the new bands are determined  by the new period $L$ and,
therefore it is proportional  to $t_1=t a^2/L^2-W$,
 where $t$ and $a$ are
the original bandwidth and interatomic distance, respectively. One can see
 that 
the value of $t_1$ is nothing but the half  distance between
the levels created by   the size quantization at zero magnetic field. 

When the magnetic field changes
each energy level 
oscillates with a single flux quantum period $\phi_0$
and  the amplitude equals to the  new bandwidth $t_1$.
Thus, the amplitude of the  oscillations
of the energy flux dependence associated with this single level
becomes  smaller as the strength of  disorder increases
\cite{Bouc}. For an infinite system the unit flux quantum period
implies a localization. In  contrast, without disorder
a single energy level oscillates  with the period equal $N \phi_0$
and the amplitude equal to the  original bandwidth $t$
\cite{Aoki}. The total persistent current is maximal when
 the  magnetic flux corresponds to the 
level crossing.
With  disorder
there appears  a finite curvature   associated with the level repulsion
(the point of the level-crossing
corresponds to an infinite curvature).
This rounding  of the intersection points 
gives  mostly  the current  decrease.
That is if the disorder increases,
the curvature of the intersection point decreases and, 
therefore,
 the suppression of the persistent current increases
 (see, for comparison, Refs.\cite{Bouc,Ambe0,Mull,Kus,Weis,Bouz}).
The total flux dependent energy
is  a sum of the single electron energies.
 The  partial current associated with a single energy level
is determined by the first derivative of its energy 
with respect to the flux. 
The total persistent current  consists of the sum of these
 partial
currents.
The dependence on the flux for each single electron energy
 becomes smoother with  disorder, 
which means that the persistent current
is strongly suppressed\cite{Weis}.
These arguments are very general and from the first view, it seems, 
that in  systems with strong electron-electron interaction we would
 expect the same picture\cite{Mull}.

However,
at a strong electron-electron interaction the pattern of
level' intersection
 is much more dense (see, for example, Refs
\cite{Kus1,Scho,Yu,KWKT,KusM/N}).
Small systems with small number of particles
can be investigated  numerically (by exact diagonalization) 
\cite{KWKT,Aoki,Chia}, while   big systems --
with the aid of the Bethe ansatz.  
In the limit of the strong electron-electron interaction
  ($U/V_F\rightarrow \infty$) in the framework of a Hubbard model the
spectrum\cite{spec} has an analytic solution of 
the universal form\cite{Kus1,Scho,Yu,KWKT,KusM/N}:
\begin{equation}
K_n = \frac{2\pi n }{L}+\frac{2 \pi }{L}( f+ \frac{ \sum_{\alpha} J_{\alpha}}{N}),
\label{momen-Hub}
\end{equation}
which is different from that
for spinless fermions: $K_n = \frac{2\pi n }{L} +\frac{2 \pi f} {L}$, by an additional
statistical (or gauge) flux 
$\phi=  {\sum_{\alpha} J_{\alpha}}/{N}$,
 where the $J_{\alpha}$ are the
quantum numbers of spinons. The total energy of the system in both cases
is determined in the same way by
the relation $E(f)=-2t \sum_n \cos K_n$ and it is equal to the holon energy
$E(f,\phi)=E_h=V_F (f-\phi)^2/L$.
 The gauge flux 
arises due to the interaction and takes the fractions
 $\phi= l/N$, 
where $l=...,-2,-1,0,1,2,...$ is any integer number.
It is related to the local  $ SU(2)$  symmetry of individual spins and
 arises due to exclusion of doubly-occupated states,
 i.e.  due to the
decoupling of the spins on the different sites and
the possibility of
individual spins  to rotate freely in  space. 
Therefore, free local spin rotations
may be 
 described by  the gauge field.
 The flux of the gauge field corresponds to the 
different spin configurations.
Since the eq.(\ref{momen-Hub}),
looks like   single particle spectrum of free spinless fermions
 with the disorder one could expect the same level repulsion  and the narrowing 
 of bands as for free electrons.

 At a fixed flux value  each new band 
is associated with a single level of a size quantization.
 With the change of magnetic flux
 the level  moves  inside the   "band" within the interval $t_1$.
 Let the  electron be located
at the bottom of the band\cite{Rem5}. 
 Then with the change of the magnetic
 flux by
 $2\pi/N$ one may tune the gauge flux by the value $-2\pi/N$. This
results in the location of an
 electron at the bottom of the band and $E_h=0$.
That is, the energy-flux dependence consists of equidistant
parabolic-like curves (see, for example,  Refs
\cite{Kus1,Scho,Yu,KWKT,KusM/N})
  and there occurs the fractional  $1/N$  -periodic Aharonov-Bohm (AB)
effect.
The current $I$ is equal to the derivative of  $E(f)$ 
with respect to the flux $f$ : $I(f)=-\partial E(f)/\partial f$ and,
therefore, the
current-flux dependence   is also a $ 1/N$ periodic function of the flux $f$
consisting of sawtooth peaks. 
 Note, however, that
 the disorder has a  little effect on the curvature
of the bottom of the energy-flux band and is acting mostly on the region
of the level intersection.
 Therefore, each of the current peaks has, approximately,   the slope of a free electron current.
The amazing fact, however, is that the weak disorder has practically 
no influence
on this  current, which, because of a small period,
has a very small amplitude $V_F/(LN)=t/U L^2$.
 The current will be suppressed only at  very strong
disorder such as $W\sim t$, when the energy-flux "band" becomes  flat.

The fractional $1/N$ period has a simple 
explanation\cite{Kus1} in a picture of  the
simultaneous, collective, cooperative motion of all $N$ electrons
on the ring which we name as
  "$N-$electron train". After one lap of the train motion
the many body wave function will get
the phase factor equal to $2\pi f N$. The gauge invariance dictates
 $fN=1$, which means that {\it the minimal periodicity of the $AB-$ effect will
be equal to $f=1/N$}. Thus, the creation of this $N-$ electron train
gives rise to a very small $1/N$ periodicity and to a nonsensitivity
of the persistent current to the weak disorder.
Thus, we offer a new mechanism of the current  based on {\it a formation of  $N-$ electron trains, which are not sensitive
to the disorder}.
Although this $N-$electron-train current is only slightly influenced
 by
disorder, its value is very small due to the big charge of the train $Q=Ne$, 
i.e. due to
the small AB period.

The first correction in the parameter $\alpha=V_F/U<<1$ 
  gives rise to an antiferromagnetic interaction
between the electron spins. Therefore, different spin configurations
will have  different  magnetic energies associated with
 the energy of the spin waves.
At a finite value $\alpha$ the spin waves velocity  is  nonzero.
Then the total energy of the Hubbard ring $E(f,\phi)$
 (let us assume, for simplicity, $N_{\uparrow}=M= {N}/{2}$) is the sum of
 the holon energy $E_h=V_F(f-\phi)^2/L$  and  the energy
of the spin waves, i.e.  
\begin{equation}
E(f,\phi) = \frac{V_F}{L} (f-\phi)^2 +\frac{V_F}{L} \alpha\mid \sin 2\pi \phi\mid 
\label{total}
\end{equation}

At zero gauge flux $\phi$ 
 the holon energy $E_h$
  increases rapidly $\sim f^2$.
However, if together with  $f$ we also change $\phi$,
 and excite the spin-waves, then the contribution of the first
term may vanish and the total energy decreases.
Although the energy in the new state with the external flux
equal $2\pi/N$  is different from the energy at  zero flux
by the spinon energy, we
still have a pronounced energy-flux oscillation with the quasi-period equal
to $2\pi /N$. The value $\phi$ is always
chosen to minimize the total energy.
One general conclusion is that  
the structure of the spin-wave excitation spectrum,
i.e. the energy dependence  on the spin wave momentum 
determines the   AB periodicity as well as the parity effect. 

For the calculation of 
the energy-flux dependence at different
ratios $M/N$ in the framework of the Bethe ansatz solution
we use the method, recently, presented in Ref.\cite{KusM/N}.
Since the $E(f)$ dependence   consists
of many segments of parabolic like curves associated with the $1/N$
oscillations (see, Figs in Ref.\cite{KusM/N}),
the current-flux dependence is calculated at each segment independently
as $I(f)=-\partial E(f)/\partial f$ and is a straight line.
The length of
 these straight lines is determined
by points of intersection of neighboring parabolic curves\cite{KusM/N}.
 The total current-flux 
dependence consists
of many parallel lines associated mostly with  single  $1/N$ oscillations.
 We have calculated the persistent current
at different ratios of $M/N$.
The most striking result, however,  is that  {\it 
the persistent current is  
perfectly ${M}/{N}$ periodic}\cite{Rem1}. 
On the  Fig.1A we present the results of our calculations
 when the number down spins
$M=101$ and $N=303$. The current is a perfect
$1/3-$ periodic   function of the flux, which was not
obvious from the analysis of
the energy-flux dependence\cite{KusM/N}. We would expect that this
new periodicity will be very approximate. It seems that the
$M/N$-fractional AB effect is much more pronounced for the current than
for the energy-flux dependence (see, for comparison, Figs in Ref.\cite{KusM/N}).
 
There are two different regions, where   the current
effectively increases with the flux and monotonously decreases. In the flux region
where the current effectively increases 
there occurs the perfect $1/N-$ or,
here, 1/303- periodic
oscillations, which are practically not resolved on the scale of
 the Figure. If the number of particles $N$ decreases such oscillations may be already seen. On the Fig.1B we present the case when $M=33$ and $N=99$. Again one can see
 perfectly  1/3-flux quantum periodic oscillations, where, however,  1/99 periodic
current oscillations are already resolved and may be clearly seen.
 The current
decreases linearly within each $1/N$ flux region. The slope of this decrease
 is equal to the slope of the free
electron current decrease. The large amplitude
of the current is determined by a large number of 
 small $1/N-$ oscillations, related
to the $N-$ particle bound  coherent state\cite{Kus1} or to the formation of
the $N-$ electron "train". One sees from  Fig.1, that it can be very 
large $I_{max}\sim 10V_F/L\sim 0.3 I_{free}$,
where  the maximum amplitude of the 
free electron current is $I_{free}=4 \pi^2 V_F/L$.
Since the weak disorder has a little effect
on the motion of this "electron" train, we expect that at the
disordered ring  the current will be  big as well.
Loosly speaking,
  since the
current is associated with this  $N-$ particle coherent state, which spreads
over the perimeter of the ring,  its
localization length  is in $N-$ times larger than for free electrons. 
 Therefore, on the Hubbard ring with the disorder the localization effects are suppressed by the strong interaction, and the current is big. 
With the interaction  also the temperature suppression of the current decreases\cite{KusTreste}.

When the value of $\alpha$ increases  the number of the $N-$
electron train excitations (or the number of the $1/N$ oscillations)
in the single AB half period decreases.
Then  there 
 the regimes with $N-2$, with  $N-6$, with  $N-8$,..., with $N-2 K$ 
oscillations  will occur,  K being any integer
(see, for comparison, the numerical
simulations in Ref.\cite{Chia}).
At   $\alpha>\alpha_c\sim 0.02$ the $N-$ electron trains disappear
and
a complete decoupling of spin and charge degrees of freedom
(Luttinger Liquid) occurs\cite{KusTreste}. However, the fractional $M/N$ oscillations do still
survive, since they are related to the $2 k_{Fs}$ excitations of the spinon
Fermi   see. On Fig.2, we show 
the 1/3 periodic oscillations of the ground state energy-flux dependence
which consists of low energy parabolic curves associated only with
charge degrees of freedom. The neighboring parabolic curves differ
by   $2 k_{Fs}$ excitations of the spinon Fermi see.
 The total spin wave excitation spectrum is presented
by short horizontal  lines on this Figure
(instead of the flux $f$  the energy  here
depends on the spin
wave momentum $k_s$, which, in our notations, is equal to $2\pi f$).
 The parabolic curves create the cusps at the flux values
$f_c\sim 0.35$ and $f_c=0$. For  other half-flux regions this dependence
must be continued with the aid of symmetrical reflections.
 One sees that in this case the persistent current
is determined by lowest parabolas or, practically, by charge degrees of freedom. 
Except $2k_{Fs}$ excitations the spinon energy is always bigger
and, therefore,
 swiches off
and does not participate in the AB effect. 
One sees on this Figure that  the $M/N$ oscillations of the current
 survive even if 
 $\alpha$ increases, which is due to the existence of the  $2k_{Fs}$ excitations.

 However, when $\alpha$ decreases,
the amplitude of the spin wave spectrum ($\sim \alpha$) decreases.
If $\alpha \leq \alpha_c$ at some flux values the spin wave energy
becomes  lower  than the cusp energy 
associated with the intersection of parabolic curves
describing charge degrees of freedom
(see, on Fig.3A, where the notations are the same as for Fig.2).
 Then, in these regions (for example, one of these regions is
 at $0.31<f<0.35$, see, Fig.3A) it is energetically
 favorable to have a spin
wave excitations, i.e. here
 the charge degrees of freedom are coupled
with the spin ones. This gives rise to the the  N-electron trains and to
 the $1/N$ oscillations of the energy-flux dependence, which
    envelope function  is determined by the spin wave spectrum.

At $U=\infty$ ($\alpha=0$) the spin waves 
 have zero velocity, and, therefore,
the spin currents $I_s$ do not exist. If $\alpha\neq 0$
the spin-current does occur
($I_s\sim \alpha$),  and it
depends on the flux of the magnetic field as a nonanalytic
function  which is  also $M/N$ periodic.
 The spin current remains  constant during each of the $1/N$ oscillations
of the electric current and decreases, step by step, with every next
$1/N$ oscillation of the electric current.
{\it The magnetic flux-dependence of the spin current  and
its periodicity with the magnetic flux }
are absolutely new effects, which are  different
 from Aharonov-Bohm effect or from Aharonov-Casher (AC) effect.
Here we have  {\it the dependence of the spin current on 
the flux of the magnetic field}, whereas
 in AB effect an electric current depends on a magnetic flux,
while in the AC effect a spin current depends on the flux of electric field.

The maximum amplitude of the spin current
 $I_{s,max}\sim V_{Fs} N_s/L\sim V_F \alpha N_s/NL$ is determined
by the spinon Fermi velocity $V_{Fs}$ and it also
depends on   the number of steps $N_s$
or 
the number of $1/N-$ periodic oscillations of the electric current 
 equal to one of the following numbers: $N$, $N-2$, $N-4$, ....
 The number of steps $N_s$ 
depends on
the ratio of the spinon and the holon Fermi velocities and decreases
when this ratio increases.
 The
amplitude of the ${1}/{N}$  oscillations of the electric current 
is proportional to $\sim{V_F}/{LN}=t/L^2$, which is a very small value
when  the ring is very large.
The surprising fact is  that the amplitude of ${M}/{N}$ 
periodic oscillations of the electric current is proportional to
 $\sim {V_F}(1-N_s/N)/L$. If $N_s<<N$, 
it is of the order of a free  electron current,
which is   much larger than the value of the spin current.

The results obtained in the framework of the Hubbard model
are very general. We have checked this picture for
  the Kondo ring (a single Kondo impurity on the thick metallic ring)
where the results are identical to those described above
and for the $SU(N)*SU(M)$ invariant models\cite{Suth,Tsve}.
At strong coupling there occurs a strong degeneracy (or a local symmetry)
 associated with the channel or spin
index. Strong local Coulomb interaction gives a freedom to
 local spin rotations or to   interchannel transitions and,
therefore, generates    a local gauge field. 
The value of  the gauge flux $\phi$, which may be not single,
are related to
  the total momentum of "spinons" or to the total
momentum of the relative charge density waves created between channels, which
may take the small fractions $\sim 1/N$.
 This small fraction of gauge  flux $\sim 1/N$, in which the "holons"
are moved, generates the "N-electron trains"
giving rise the fine structure of the AB effect
(the fractional $1/N$ and ${M}/{N}$ periodic oscillations)
 and making the current  
insensitive to the disorder. 
Of course, for the real interaction, which is not infinite, this degeneracy
is lifted. This means that the creation of the gauge flux costs some energy,
which is, however, small if the parameter $V_F/E_c$ is small.
The  picture is valid for any multichannel or 2D ring, where 
the interchannel interaction is so strong that the gauge field
arises due to the
 interchannel transitions (see, for comparison, Ref.\cite{Aoki}).

  Whether or not this "train" effect can
explain the discrepancy between experiments \cite{22,21,33}
and existing theories
is still unclear. The detection of any described fractional periodicities
may resolve this question and prove the existence of the "train"
excitations on the quantum rings.
Since the distribution of particles over the channels is, probably, well
controllable, it should be a realizable experiment
to see the predicted
fractional periods. One may also expect the described persistent current with fractional periods $p/q$ in higher magnetic field, where electrons are partially
polarized.
 
I thank Markus Buttiker, C.A.Stafford, D.E. Khmel'nitzkii, 
D.V. Khveshchenko, Alan Luther, E. Mucciollo for nice conversations.
 
{\bf REFERENCES}

permanent address:  {\it Department of  Physics,}\\
  {\it Loughboro' University, Loughboro', Leicestershire, LE11 3TU, UK} \\


{\large \bf Figure Captions}\\

\bigskip

{\bf Fig.1}  The behavior of the persistent current as a function of
flux  $f$ within the half of fundamental flux quantum:
 A) for 303 electrons at the values $L=20000$ and $U=10$ and there are
$M= 101$ particles with up-spin;
B)  for 99 electrons at the values $L=10000$, $U=10$ and $M=33$.
On the slope where  the current  increases, there are perfect 1/303 (for A) and
1/99 (for B) flux quantum periodic oscillations.
In both cases the maximal current amplitude is equal to $\sim 10 V_F/L$ and 
its main periodicity  is equal to 1/3 in units of elementary flux quantum.

{\bf Fig.2}  The behavior of the ground state
 energy  as a function of
flux  $f$ (the lower parabolic-cusp curve)
and the spinon energy (indicated by the short horizontal lines)
for 303 electrons with the 101 up spin electrons
at the values $L=1000$ and $U=10$ in the region
 within the half of fundamental flux quantum.
The energy is expressed in 
the units $t  10^2$. The zero energy corresponds to $-520.0 t$.

\bigskip

{\bf Fig.3}  A) The behavior of the holon
 energy   (the  parabolic-cusp curve) and the spinon energy,
indicated by short horizontal lines
 as a function of
flux  $f$ ($\Phi$) for 303 electrons ($M=101$) at the values $L=20000$ and $U=10$ in the region within the half of fundamental flux quantum.
 The zero energy corresponds to $-605.771443 t$.
The energy is expressed in 
the units $t  10^7$. B) The behavior of the ground state
 energy  as a function of
flux  $f$.

\bigskip


\begin{references}
\bibitem{spec}
Although the form of the spectrum is the same for several
models, without loosing  the generality, we  speak  about 
the Hubbard model, only.
\bibitem{Rem5}  For simplicity we assume that there is odd number of electrons
on the ring. In this case at small values of the flux the current is 
determined by electron located on the bottom of the band.
The other situations with different parity  are considered analogously.
\bibitem {Rem1} There is also single flux quantum periodic oscillations
of the persistent current,
which amplitude is, however, very small when $M$ of the order $N/2$ and 
which amplitude increases when $M$ decreases.
\end{references}
\end{document}